\documentclass[12pt,preprint]{aastex}

\slugcomment{Published in ApJ}

\shorttitle{Emergence of Kink-Unstable Magnetic Flux Tube}
\shortauthors{Takasao et al.}

\usepackage{bm}
\usepackage{amsmath}
\usepackage{color}

\begin{document}


\title{NUMERICAL STUDY ON EMERGENCE OF KINKED FLUX TUBE FOR UNDERSTANDING OF POSSIBLE ORIGIN OF $\delta$-SPOT REGIONS}


\author{Shinsuke Takasao\altaffilmark{1}}
\email{takasao@kwasan.kyoto-u.ac.jp}

\author{Yuhong Fan\altaffilmark{2}}
\author{Mark C. M. Cheung\altaffilmark{3}}

\and

\author{Kazunari Shibata\altaffilmark{1,4}}


\altaffiltext{1}{Kwasan and Hida Observatories, Kyoto University, Yamashina, Kyoto 607-8471, Japan}

\altaffiltext{2}{HAO, National Center for Atmospheric Research, P.O. Box 3000, Boulder, CO 80307, USA}

\altaffiltext{3}{Lockheed Martin Solar and Astrophysics Laboratory, 3251 Hanover Street Bldg. 252, Palo Alto, CA 94304, USA}

\altaffiltext{4}{Unit of Synergetic Studies for Space, Kyoto University, Yamashina, Kyoto 607-8471, Japan}

\begin{abstract}
We carried out a magnetohydrodynamics simulation where a subsurface twisted kink-unstable flux tube emerges from the solar interior to the corona. Unlike the previous expectations based on the bodily emergence of a knotted tube, we found that the kinked tube can spontaneously form a complex quadrupole structure at the photosphere. Due to the development of the kink instability before the emergence, the magnetic twist at the kinked apex of the tube is greatly reduced, although the other parts of the tube is still strongly twisted. This leads to the formation of a complex quadrupole structure: a pair of the coherent, strongly twisted spots and a narrow complex bipolar pair between it. The quadrupole is formed by the submergence of a portion of emerged magnetic fields. This result is relevant for understanding of the origin of the complex multipolar $\delta$-spot regions that have a strong magnetic shear and emerge with polarity orientations not following Hale-Nicholson and Joy Laws.
\end{abstract}

\keywords{Sun: corona --- Sun: magnetic field --- Sun: interior}

\section{INTRODUCTION}
Magnetic flux emergence from the solar interior to the overlying atmosphere is responsible for the formation of active regions \citep{parker1979,fan2009,cheung2014}, and therefore it is the driver of various explosive phenomena \citep{shibata2011,takasao2013}. The free energy buildup in the corona during flux emergence, which is essential for the eruptive events, has been extensively studied by many authors \citep[e.g.][]{magara2003,manchester2004,fang2012,cheung2012}.

It has been known that certain photospheric configurations of active regions are closely related to the flare activity \citep[e.g.][]{kurokawa1989,shi1994}. The so-called "$\delta$-spot regions" are known to be among the most flare-productive active regions \citep{sammis2000}, in which sunspot umbrae of opposite magnetic polarities are pressed together in a common penumbra. The population of the $\delta$-spot regions is small, but they are the main source regions of the large flares (According to \citet{sammis2000}, more than 80\% of X-class flares occur in $\delta$-spot regions). In this sense, $\delta$-spot regions have a significant impact on the space weather.

Observations show that a fraction of the $\delta$-spot regions have strong magnetic shear along the polarity inversion line and emerge with polarity orientations not following the Hale-Nicholson and Joy Laws (hereafter, Joy's law) obeyed by the majority of active regions \citep{zirin1987}. It is speculated from observations that they are formed as a result of the emergence of current-carrying (twisted) flux tubes \citep[e.g.][]{leka1996}, which implies that a considerable amount of the free energy is stored in such $\delta$-spot regions. In addition, they tend to appear with complex multipolar spots \citep{zirin1973,ishii1998}. Multipolar regions are generally more favorable for the ejections than bipolar regions \citep{antiochos1998}, because smaller energy is enough for plasma to escape from the closed magnetic field region (the number of the field lines that plasma has to stretch is much smaller than that for the bipolar systems). Therefore, their strong magnetic shear and multipolar structure are important characteristics of the flare-active $\delta$-spot regions.

It is suggested that $\delta$-spot regions are formed as a result of the emergence of the twisted flux tubes with an abnormal structure. \citet{tanaka1991} analyzed the time evolution of the August 1972 $\delta$-spot region in detail. He found that the behavior of this region can be explained by the emergence of a twisted flux tube with a knotted structure, and the twist and writhe of the flux tube having the same sign. \citet{kurokawa2002} also reported an event which is likely related to the emergence of a knotted flux tube.

If a flux tube is sufficiently twisted, then it can be subject to the kink instability below the photosphere, where the kink instability is a magnetohydrodynamic (MHD) instability of a highly-twisted flux tube. When the kink instability sets in, the field-line twist about the axis will be converted to the writhe of the axis as a result of helicity conservation (the twist around the axis will be reduced), and therefore the tube will be knotted or kinked \citep{moffatt1992}. It should be noted that the twist and writhe in this case have the same sign, where a positive/negative twist (writhe) is defined as right/left-handed twist (writhe).

It is believed that the sign of the tilt angle of an active region with respect to the solar equator can be used as a proxy for the sign of the writhe. Therefore, investigating the relation of the sign for the twist and tilt from observations is an important step to check the possibility of the kink instability scenario. Statistical studies of the correlation between active region twist and tilt angles have been performed \citep{tian2003,holder2004,tian2005}. \citet{holder2004} found a significant correlation between active region twist and tilt angles by analyzing 368 active regions, where the correlation between them denotes that twist and writhe have the same sign. They identified that the correlation is mainly contributed by those active regions (174 of 368 regions) that deviate significantly from Joy's law. \citet{tian2005} obtained a similar result by surveying 104 $\delta$-spot regions. \citet{takizawa2015} also confirmed this result by looking at the birth phase of the 31 flare-productive $\delta$-spot regions in the cycle~23. Detailed case studies \citep[e.g.][]{tanaka1991, lopez-fuentes2003} have shown that a non-negligible fraction of these active regions seems formed by the emergence of kinked flux tubes with the same sign for twist and tilt. \citet{nandy2006} found the tendency that $\delta$-spot regions have stronger twist than others, which may support the idea of the kink instability scenario.

There are several theoretical investigations of kink unstable flux tubes in the context of the formation of flare-productive active regions \citep{linton1996,matsumoto1998,linton1999,fan1999}. \citet{matsumoto1998} performed the first 3D MHD simulations of the emergence of kinked flux tubes into the corona, and they argued that the emergence of kinked flux tubes can explain the formation of a sequence of (strongly sheared) S-shaped active regions. \citet{linton1999} studied basic properties of a kink unstable flux tube with high plasma-$\beta$, but neglected gravitation, where the plasma-$\beta$ is defined as the ratio of the gas pressure to the magnetic pressure. \citet{fan1999} performed several sets of MHD simulations of the rise of kink unstable flux tubes in the interior of the Sun. They found that the kinking motion is promoted by the gravitational stratification, and that a kinked tube has strong shear along the polarity inversion line (PIL) in the buckled part. Since the models by \citet{linton1996} and \citet{fan1999} are confined to the solar interior, it remains unclear if the kinked tube can emerge into the atmosphere to produce the observational characteristics of $\delta$-spot regions. Moreover, another important observational feature is that many $\delta$-spot regions contain multipolar spots \citep[e.g.][]{zirin1973}, which is not explained by the kinked tube model of \citet{linton1996} and \citet{fan1999}.

Other mechanisms to form $\delta$-spot regions have also been considered. For instance, \citet{toriumi2014} and \citet{fang2015} performed MHD simulations of the emergence of a single twisted (kink stable) flux tube with two buoyant segments, and they successfully obtained $\delta$-spot-like regions as a result of the collision of the non-paired spots. \citet{toriumi2014} showed that the sheared polarity inversion line does not form when the pair of photospheric bipoles are due to the emergence of two adjacent (but unconnected) flux tubes. Considering their result, the multipolar spots should be magnetically connected below the surface to keep the active region compact.

In this paper, we present the results of an MHD simulation in which a subsurface twisted kink-unstable flux tube emerges from the solar interior into the corona. From this simulation we found that a complex quadrupole structure is spontaneously formed as a result of the emergence of a single kinked flux tube. The remainder of the paper is structured as follows. Section~\ref{sec:setup} describes the numerical setup of our simulation. In Section~\ref{sec:results}, we show numerical results of the emergence of a kinked flux tube, and briefly compare our simulation with observations. Finally, in Section~\ref{sec:discussion}, we discuss our findings through comparison with previous studies, and also discuss the implications of this work for understanding of the formation of $\delta$-spot regions.

\section{NUMERICAL SETUP}\label{sec:setup}
\subsection{Basic Equations}
We solved the MHD equations in the following form:
\begin{align}
\frac{\partial \rho}{\partial t} &=- \nabla \cdot(\rho \bm{v}), \\
\frac{\partial \rho \bm{v}}{\partial t} &= - \nabla \cdot \left[ \rho \bm{vv} 
+ \left( p+\frac{|\boldmath{B}|^2}{8\pi}\right) \underline{\bm{1}} - \frac{\bm{BB}}{4\pi} \right] + \rho\bm{g},\\
\frac{\partial \bm{B}}{\partial t} &=- \nabla\cdot (\bm{vB} - \bm{Bv}) -\nabla \times (\eta \nabla \times \bm{B})\label{eq:ind},\\
\frac{\partial e}{\partial t} &=- \nabla \cdot \left[ \bm{v} \left( e + p+\frac{|\boldmath{B}|^2}{8\pi}\right) - \frac{1}{4\pi}\bm{B}(\bm{v}\cdot \bm{B}) \right] + \frac{1}{4\pi}\nabla\cdot(\bm{B}\times\eta\nabla \times \bm{B})+ \rho(\bm{g} \cdot \bm{v}),\\
e &= \frac{p}{\gamma-1}  + \frac{\rho |\bm{v}|^2}{2} + \frac{|\bm{B}|^2}{8\pi}, &
\end{align}
where $\rho$ is the density, $\bm{v}$ is the velocity vector, $\bm{B}$ is the magnetic field vector, $e$ is the total energy density, and $p$ is the gas pressure. $\gamma$ is the specific heat ratio (here $5/3$). The equation of state for the ideal gas is used. The gravity $\bm{g}=-g_0\hat{\bm{z}}$ is a constant vector and the non-dimensional form is given by $\bm{g}=(0,0,-1/\gamma)$. The normalization units of our simulations are summarized in Table~\ref{tab:units}. In the expanding magnetic field in the corona, the plasma $\beta$ can become very small, where the calculated gas pressure can be negative due to numerical errors. To avoid the numerical instability, the energy equation in the low-beta ($\beta<0.01$) regions is replaced by the following equation:
\begin{align}
\frac{\partial e_{int}}{\partial t} =- \nabla \cdot (\bm{v}e_{int}) -(\gamma -1 )e_{int}\nabla \cdot \bm{v},
\end{align}
where $e_{int}=p/(\gamma-1)$. In Equation~\ref{eq:ind}, $\eta$ is the magnetic diffusivity, and an anomalous resistivity model is assumed:
\begin{eqnarray}
\eta=\left\{ \begin{array}{ll}
0 & (J < J_c \hspace{2mm}{\rm or}\hspace{2mm}  \rho > \rho_c)\\
\eta_0(J/J_c -1) & (J\ge J_c \hspace{2mm}\&\hspace{2mm} \rho < \rho_c), \\
\end{array} \right.
\end{eqnarray}
where $\eta_0=0.1$, $J_c=0.1$ and $\rho_c=0.1$. Density is normalized such that $\rho=1$ corresponds to the density at the photospheric base (see also Table~\ref{tab:units}). Thus, the resistivity works only in the region where the current density is strong above the photosphere.

The numerical scheme is based on \citet{vog2005}: Spatial derivatives are calculated by the 4th-order central differences and temporal derivatives are integrated by a 4-step Runge-Kutta scheme. We also adopt the artificial diffusion term described in \citet{rem2009} to stabilize the numerical calculation. Errors caused in $\nabla\cdot \bm{B}$ are controlled by using an iterative hyperbolic divergence cleaning method, where the cleaning technique is based on the method described in \citet{ded2002}.

\subsection{Initial and Boundary Conditions}
\begin{figure}
\epsscale{.80}
\plotone{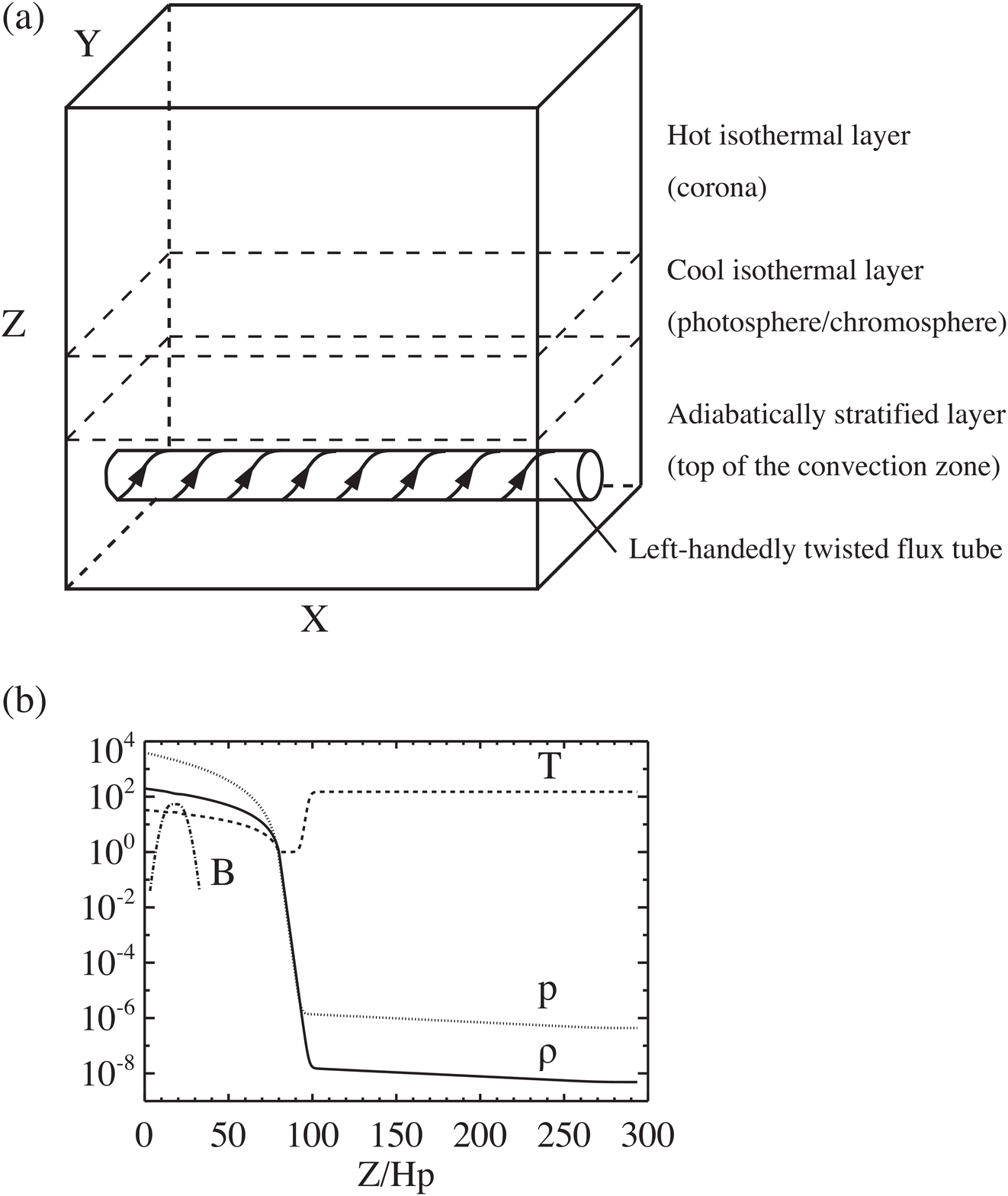}
\caption{Initial setup of the simulation. (a) Schematic diagram of the model atmosphere and initial flux tube. (b) Vertical distributions of the density $\rho$ (solid), pressure $p$ (dotted), temperature $T$ (dashed), and magnetic field strength $B$ (dashed dotted line) in the initial condition. See Table~\ref{tab:units} for the normalization units. \label{fig:ic}}
\end{figure}

The initial background atmosphere consists of three regions: an adiabatically stratified static atmosphere (representing the convection zone), a cool isothermal layer (photosphere/chromosphere), and a hot isothermal layer (corona) (see Figure~\ref{fig:ic} (a)). The initial distribution of temperature is given as
\begin{align}
T(z) = T_{ph} - \left|\frac{dT}{dz} \right|_{ad} z
\end{align}
for the convection zone ($0\le z \le z_{ph}$), and
\begin{align}
T(z) = T_{ph} + (T_{co}-T_{ph})\left\{ \frac{1}{2}\left[ \tanh\left( \frac{z-z_{tr}}{w_{tr}}\right) +1 \right]\right\},
\end{align}
for the upper atmosphere ($z\le z_{ph}$), where $T_{ph}$ and $T_{cr}$ are the temperatures in the photosphere and the corona, and are set to $T_0$ and $150T_0$, respectively. $|dT/dz|_{ad}\equiv (\gamma-1)/\gamma$ is the adiabatic temperature gradient. The photospheric height is $z_{ph}=80H_{p0}$, and the transition region between the chromosphere and the corona is located at $z_{tr}=z_{ph}+18H_{p0}=98H_{p0}$. The width of the transition $w_{tr}$ is set to $2H_{p0}$. 
A magnetic flux tube is initially located below the photosphere (the depth is $\sim 10$~Mm). The longitudinal and azimuthal components of the flux tube are respectively given by
\begin{align}
B_x(r) & = B_{\rm tube,0}\exp{\left ( -\frac{r^2}{R_{\rm tube}^2}\right)},\\
B_\theta(r) & = \alpha \frac{r}{R_{\rm tube}}B_x(r),
\end{align}
where $r = \left[ y^2 + (z-z_{\rm tube})^2 \right]^{1/2}$, $z_{\rm tube}=18H_{p0}$, and $B_{\rm tube,0}=53B_0$. $R_{\rm tube}=5H_{p0}$ is the radius of the tube. These profiles are truncated at $r=3R_{{\rm tube}}$. The total magnetic flux of $3\times 10^{20}$~Mx is assumed. The plasma $\beta$ at the center of the initial tube is approximately 20. $\alpha$ is a measure of the twist of the flux tube, and is set to $-1.5$ (negative sign denotes the left-handed twist). Note that $\alpha$ is defined as a non-dimensional constant here. $\alpha/R_{\rm tube}$ gives the rate of field line rotation per unit length along the tube. The absolute value of $\alpha$ is larger than the critical value for the kink instability, 1, and therefore the flux tube is initially kink-unstable. The vertical distributions of the density, pressure, temperature, and magnetic field strength in the initial condition are shown in Figure~\ref{fig:ic} (b).  

The simulation domain is $(-124H_{p0},-124H_{p0},0)\le (x,y,z) \le (124H_{p0},124H_{p0},294H_{p0})$ resolved by a $600\times 600 \times 740$ grid. In the $x$ and $y$-directions, the mesh size is $\Delta x = \Delta y = 0.25H_{p0}$ within $|x|,|y|<30H_{p0}$ and gradually increases up to 0.8$H_{p0}$ for $|x|,|y|>30H_{p0}$. In the $z$-direction, the mesh size is $\Delta z=0.24H_{p0}$ within $0\le z \le z_{ph}+40H_{p0}$, and for $z > z_{ph}+40H_{p0}$, it gradually increases up to 1.6$H_{p0}$. To facilitate investigation of the coupling between the kink and Parker instabilities, we take a domain size large enough to allow the latter to develop. The critical wavelength for it is $\sim9$ times the local pressure scale height, $\sim230H_{p0}$, which is comparable to or smaller than the calculation domain size in the $x$-direction, 248$H_{p0}$.

We make the flux tube buoyant by introducing a small perturbation to the density in the flux tube. The perturbation is described as
\begin{align}
\rho &= \rho_0 \left[ 1-a(x)\right],\\
a(x) &= \frac{1}{\beta}\left[ (1+\epsilon)\exp(-x^2/\lambda^2)-\epsilon \right],
\end{align}
where $\rho_0$ is the unperturbed density, $\epsilon=0.2$ and $\lambda=15H_{p0}$. We have checked that the evolution of the rising flux tube is not sensitive to the wavelength of the initial perturbation.

The boundaries in the $x$-direction (the direction of the initial flux tube axis) are assumed to be periodic. The boundaries in the $y$ and $z$-directions are assumed to be a perfectly conducting wall and is non-penetrating. Many previous studies adopted a fee boundary condition for the top boundary, but we applied the non-penetrating boundary for it to avoid numerical instabilities arising from the use of a free boundary condition. The top boundary is located at a much higher position ($z=294H_{p0}$) than the top of the emerging coronal loops ($z\sim 200H_{p0}$), and therefore the top boundary will not significantly affect the evolution of the flux emergence process.

\section{NUMERICAL RESULTS}\label{sec:results}
\subsection{Evolution of Flux Tube in the Solar Interior}
\begin{figure}
\epsscale{.80}
\plotone{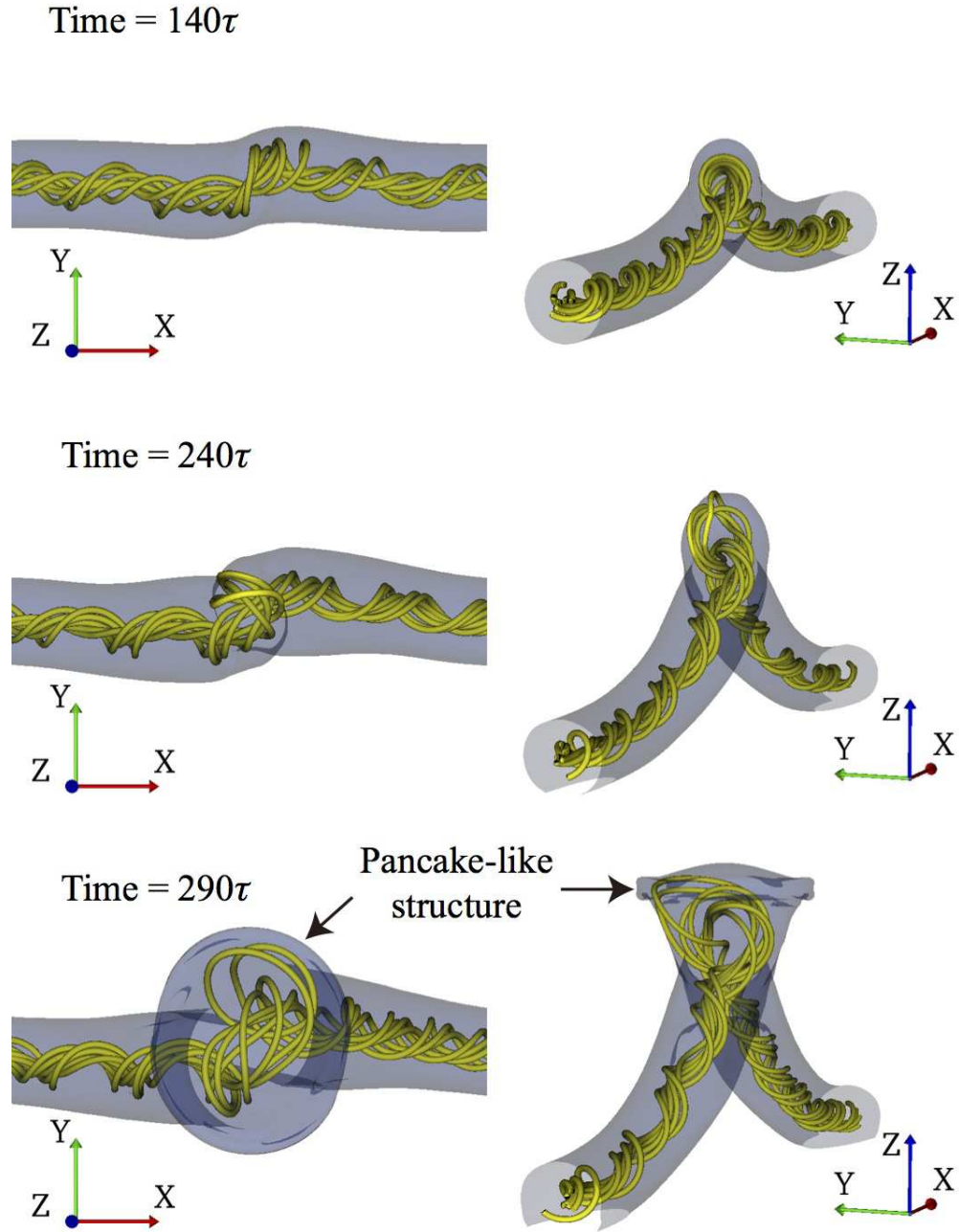}
\caption{The rise of the flux tube in the solar interior. The yellow lines denote the selected magnetic field lines, and the blue surfaces indicate the isosurface where $B=2B_0$. \label{fig:rising_tube}}
\end{figure}

The rise of the flux tube with a single buoyant segment is shown in Figure~\ref{fig:rising_tube}. Since the tube is initially kink-unstable, the knotted structure develops during the rise, as in the previous study by \citet{fan1999}. In this study, the box size in the $x$-direction is $\sim 9$ times larger than the pressure scale height at the initial tube axis, whereas they are similar in \citet{fan1999}. With the larger domain size, the rising tube develops a $\Omega$-shape with a central kinked part. 

\begin{figure}
\epsscale{.80}
\plotone{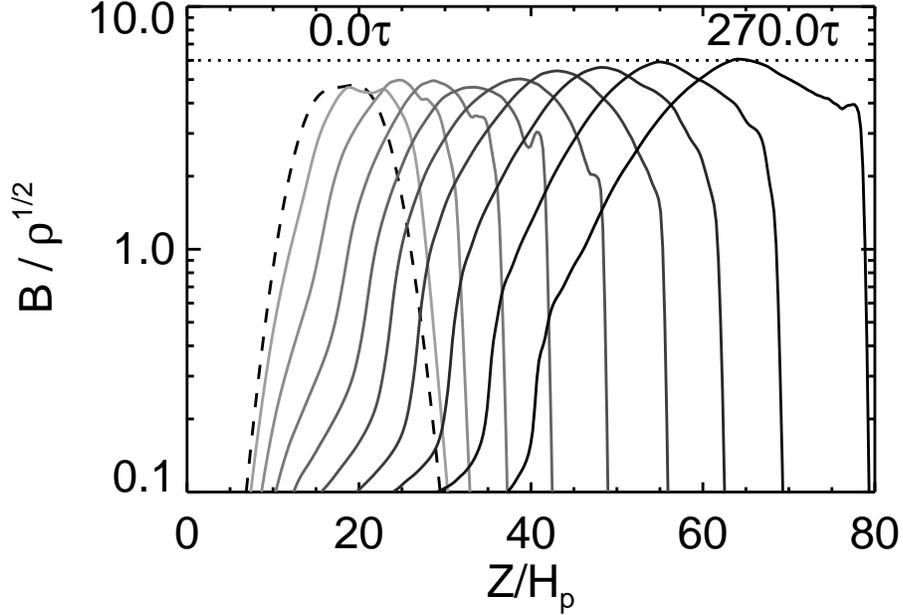}
\caption{Temporal evolution of the ratio of the magnetic field strength $B$ to the square root of the density $\rho^{1/2}$. The ratio is measured along the $z$-axis crossing the point $(x,y)=(0,0)$. The dashed line denotes the initial profile. The solid lines with different colors indicate the profiles at different times. The time interval between successive plots is $30\tau$. \label{fig:b_rosq}}
\end{figure}

To examine the deformation and expansion of the apex of the rising tube, we looked at the relation between the plasma density and the magnetic field strength of the most buoyant part. We confirmed that the magnetic field strength of the most buoyant part can be well described by  $B\propto \rho^{1/2}$ (see Figure~\ref{fig:b_rosq}), which means that the flux tube experiences a strong horizontal expansion during its rise to the surface \citep{cheung2010}. The strong horizontal expansion leads to the formation of a pancake-like (flat, horizontally extended) magnetic distribution below the surface (see Figure~\ref{fig:rising_tube}). Because of the deformation by the kinking and the strong horizontal expansion, it is difficult to deduce the resulting photospheric and coronal magnetic structures only from the magnetic structure of the rising kinked flux tube in the interior of the Sun.

\subsection{Formation of Complex Quadrupole Structure at Photospheric Level}
\begin{figure}
\epsscale{.60}
\plotone{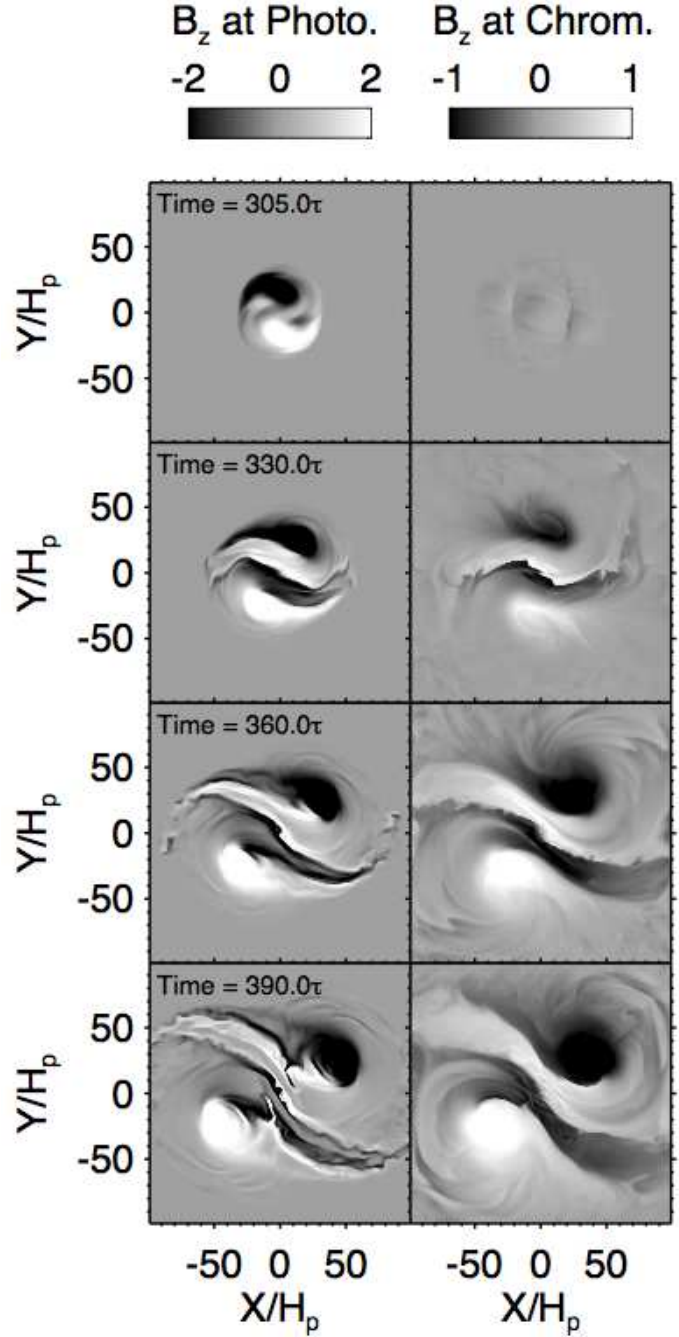}
\caption{Temporal evolution of the photospheric (Left) and chromospheric (Bottom) line-of-sight magnetic fields $B_z$. The heights of the photospheric and chromospheric planes are $80H_{p0}$ and $96H_{p0}$, respectively. $H_{p0}$=170~km~s$^{-1}$, and $\tau$=25~s. \label{fig:bz_evo}}
\end{figure}

The evolution of the vertical components of the photospheric ($z=80H_{p0}$) and chromospheric ($z=96H_{p0}$) magnetic field, $B_z$, is displayed in Figure~\ref{fig:bz_evo}. At the beginning of the emergence ($t=305\tau$), a pair of opposite polarity spots (we call them the ``main pair") appear with a large inclination with respect to the initial axis direction ($x$-direction). As time progresses, another pair of strong magnetic regions appear at the middle of the main pair of spots (called the ``middle pair"). Later, the structure of the middle pair becomes disordered, although the main spots show a coherent structure and strong twist. The main pair and middle pair have the maximum field strength of $\sim10B_0$ and $\sim3$-$4B_0$, respectively.

\begin{figure}
\epsscale{.60}
\plotone{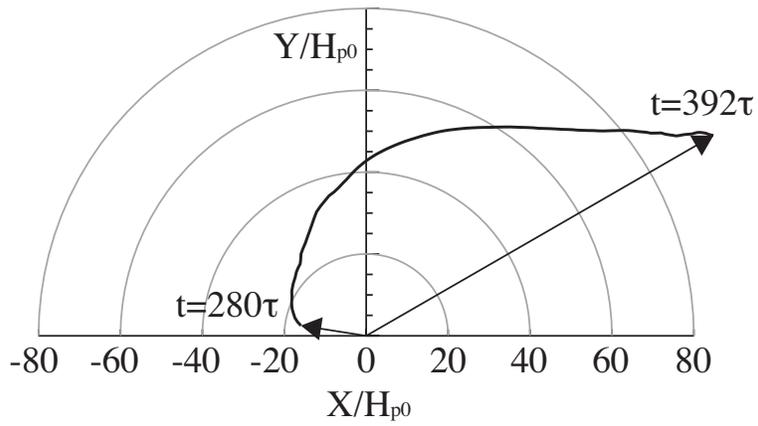}
\caption{Temporal evolution of the direction and distance from the centroid of the positive polarity region to the centroid of the negative polarity region. The centroids are defined by Equation~\ref{eq:centroid}.  $H_{p0}$=170~km~s$^{-1}$, and $\tau$=25~s. \label{fig:relative_pos}}
\end{figure}

In order to describe the motion of this quadrupole region, we measured the flux-weighted centroids of positive and negative polarities $(x_{\pm},y_{\pm})$, where
\begin{align}
(x_{\pm},y_{\pm}) = \left( \frac{\Sigma x B_{z,\pm}}{\Sigma B_{z,\pm}},\frac{\Sigma y B_{z,\pm}}{\Sigma B_{z,\pm}}\right).\label{eq:centroid}
\end{align}
Relative motion of the positive and negative polarities at the photosphere is shown in Figure~\ref{fig:relative_pos}. The centroid position for each polarity is computed including all the parts of that polarity (i.e. include both the main pair and the middle pair). The tilt angle of this region, measured from the positive $x$-direction, is large at the beginning of the emergence (note that the direction of the initial tube axis is in the $x$-axis). However, it becomes smaller later. This change may be regarded as a clockwise motion of this region, as predicted by the emergence of a knotted tube with a left-handed twist \citep{tanaka1991,linton1999}. We note that the distance between the flux-weighted centers of the opposite polarities increases as time progresses.

It should be noted that a complex quadrupole structure is formed by the emergence of a kinked tube with a single buoyant segment, not with two buoyant segments assumed in the previous studies by \citet{toriumi2014} and \citet{fang2015}. The narrow middle pair becomes prominent well after the emergence of the main pair, and its structure gets disordered, but the chromospheric $B_z$ maps in Figure~\ref{fig:bz_evo} show a more smooth quadrupole distribution than the photospheric distribution.

\begin{figure}
\epsscale{.80}
\plotone{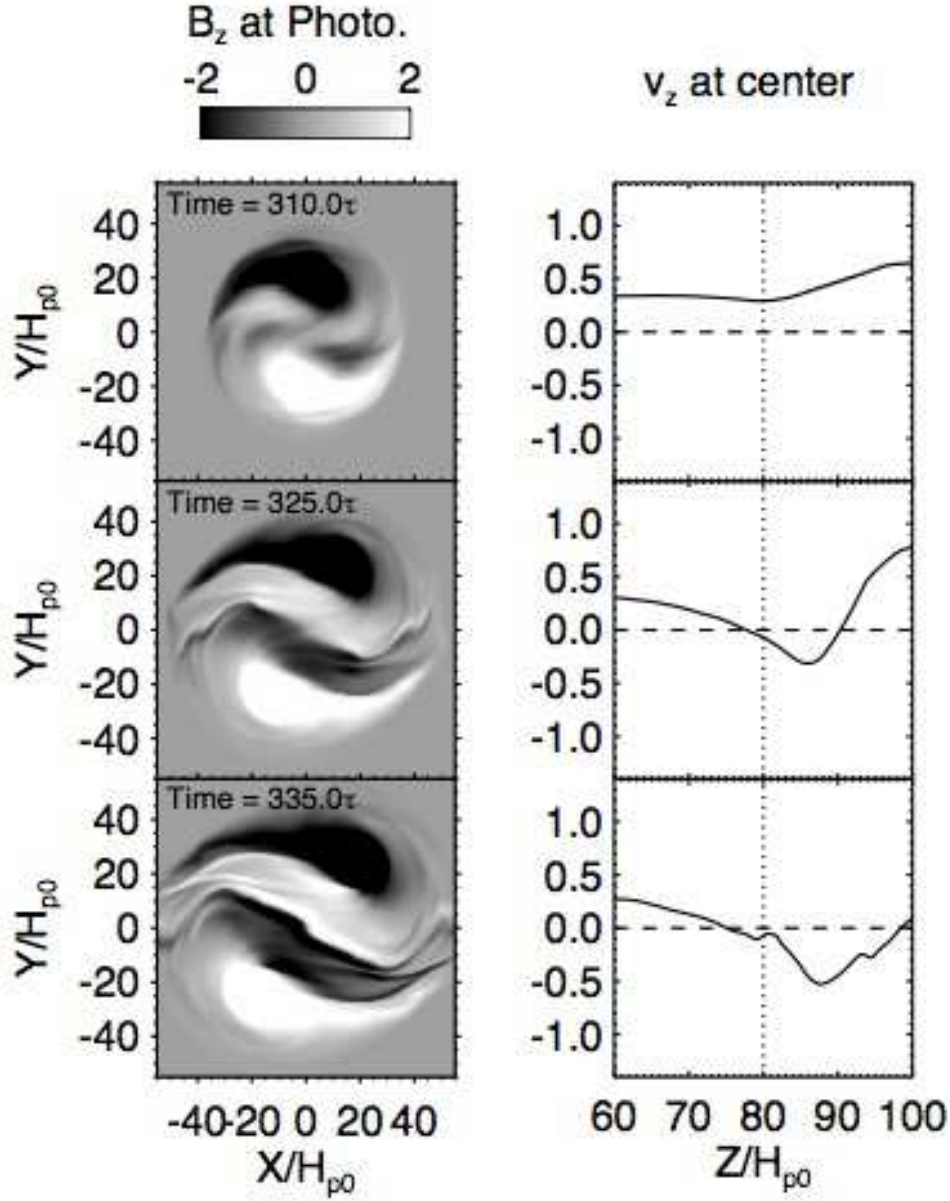}
\caption{Time evolution of the photospheric ($z=80H_{p0}$) line-of-sight magnetic field $B_z$ (Left) and the vertical velocity component $v_z$ measured at the center $(x,y)=(0,0)$ (Right). In the right panels, the horizontal dashed lines denote the $v_z=0$, and the vertical dotted lines indicate the photospheric height. Note that the sinking motion is seen when the middle pair is formed. \label{fig:vz_check}}
\end{figure}

We investigated the formation of the middle pair. Figure~\ref{fig:vz_check} shows the temporal evolution of the vertical velocity $v_z$ at the center $(x,y)=(0,0)$ during the formation. Magnetic field is rising in the early phase ($t=310\tau$), but later $v_z$ becomes negative, which means the submergence of emerged magnetic fields. Figure~\ref{fig:sinking3D} (a) displays the 3D magnetic field evolution, where it is shown that the middle pair is formed as emerged magnetic fields submerge. For this reason, the middle pair appears after the formation of the main pair. To clarify the cause of the submergence, we measured the vertical forces at the center. Figure~\ref{fig:sinking3D} (b) indicates that the sum of the upward Lorentz force and pressure gradient becomes weaker than the downward gravitational force, which means that the submergence is caused by the downward motion of the heavy material.

\subsection{Evolution of Coronal Magnetic Field}
Figure~\ref{fig:2dcut} (a-c) illustrate the evolution of the magnetic field and density distribution in the $y$-$z$ plane at $x=0$. Two magnetic flux concentrations are located just below the photosphere at the beginning of the emergence ($t=300\tau$) with a single arcade field emerging into the atmosphere. In this model, as in many previous numerical models of emerging flux, the expansion of the magnetic arcade into the atmosphere is enabled by a magnetic Rayleigh-Taylor instability \citep[e.g.][]{shibata1989,matsumoto1993,archontis2004}. As the expanding arcade plows through the atmosphere, plasma is compressed above the arcade, leading to a top heavy layer (see Figure~\ref{fig:2dcut} panel~d). Due to the continued emergence of the two strong flux concentrations, and the comparatively weaker twist at the middle between the two concentrations, the top heavy layer is driven to accumulate plasma at a central dip (see Figure~\ref{fig:2dcut} panels d and e). The continued plasma accumulation results in the submergence of the heavy plasma at the dip and the formation of the two adjacent magnetic arcades.

A snapshot of the 3D coronal magnetic structure is shown in Figure~\ref{fig:coronal_field}. The current sheet indicated in Figure~\ref{fig:2dcut} (c) is located between the blue and yellow magnetic arcades, where magnetic reconnection takes place. Two sets of the new loops colored purple and white are interpreted as reconnected field lines (Figure~\ref{fig:coronal_field} (a-d)). Looking at the reconnection site, we see the reconnection angle (the angle between the merging field lines) is not 180~degree (i.e. not perfectly antiparallel), and merging field lines have a large guide field (panel~(e)). As shown in Figure~\ref{fig:coronal_field}(d) and (f), the lower new arcade (purple lines) is almost parallel to the polarity inversion line, and is connected to the middle pair. The upper new arcade (white lines), which was ejected upward from the reconnection site, connects the far ends of the two magnetic arcades (yellow and blue).

We also note the magnetic connectivity above the photosphere shown in Figure~\ref{fig:coronal_field}. A fraction of the magnetic field of the main spots is connected to the middle pair. The other magnetic field connects the two main spots. A portion of the magnetic field connecting the main spots is the reconnected field (white lines in Figure~\ref{fig:coronal_field}).

We have seen that in Figure~\ref{fig:coronal_field}, both the yellow and blue arcades on two sides of the polarity inversion line (PIL) as well as the (purple) reconnected loops connecting the middle pair are highly sheared along the PIL of the middle pair. When we look at the photospheric motion, we find the vortical or rotational motion within each of the two main polarities (Figure~\ref{fig:wz_photo} (a)). Figure~\ref{fig:wz_photo} (b) shows the temporal evolution of the vertical vorticity $\omega_z$ averaged over the area where $|B_z|$ is above 75\% of the peak $|B_z|$ value. From the figure, we find that the counterclockwise vortical motion becomes prominent at time $t\sim320\tau$, and persists throughout the subsequent evolution. To illustrate the development of magnetic shear, we show the temporal evolution of the vector magnetogram of the middle pair at the photosphere and middle chromosphere in Figure~\ref{fig:bz_bhvec_multi_height}. We can clearly observe that the horizontal magnetic field becomes more parallel to the PIL at those heights as time progresses. Note that the horizontal magnetic field strength is small just on the PIL. This is because the middle pair is formed as a result of the submergence, not the emergence. As investigated by \citet{fan2009}, the vortical motion (and the horizontal motion) of the main polarities builds-up magnetic shear in this simulation. We also note that the magnetic shear near the PIL is not due to the direct emergence of a sheared field.

\subsection{Magnetic and Flow Structures in and near Middle Pair}
The middle pair is pushed and confined in a narrow region after the formation (Figure~\ref{fig:bz_evo}). To understand the confinement mechanism, we looked at the flow structure and the Lorentz force at the photosphere. The top panel of Figure~\ref{fig:bz_vhvec_lorentz} shows horizontal velocity vector on the magnetogram, and we can see the converging flow to the PIL. This converging flow is a natural consequence of the development (expansion) of the two magnetic arcades. The bottom panel of Figure~\ref{fig:bz_vhvec_lorentz} displays the Lorentz force vector, and we can find that the horizontal Lorentz force is pushing the two polarities together. We checked the pressure balance across the PIL of the middle pair. Figure~\ref{fig:pressure_balance_PIL} displays the profiles of the total pressure, gas pressure, magnetic pressure, and dynamic pressure across the PIL. The total pressure is almost constant across the PIL, which will explain the persistent existence of the middle pair. We can see that the high gas pressure region is supported by the magnetic pressure. We also found that the magnetic pressure has its peaks at the edges of the high pressure region because of the dynamic pressure of the converging flows.

The horizontal magnetic field is important for producing the large magnetic pressure near the PIL. As shown in Figure~\ref{fig:wz_photo}, the main polarities show the vortical motion to shear the field. Figure~\ref{fig:shear_evo3D} (a) displays the development of the horizontal field along the PIL (see also Figure~\ref{fig:bz_bhvec_multi_height}). Figure~\ref{fig:shear_evo3D} (b) shows the ratio of the horizontal field $B_h=\sqrt{B_x^2+B_y^2}$  to the vertical magnetic field $B_z$. It is found that the horizontal field is much stronger than the vertical field outside the middle pair and the PIL is sandwiched in the strong horizontal field regions. Therefore, the development of the magnetic shear along the PIL by the vortical motion of the main polarities is a key to confine the narrow middle pair.

We also found persistent fast shear flows along the PIL (see the top panel of Figure~\ref{fig:bz_vhvec_lorentz}). The maximum velocity is about $2C_{s0}$, which is supersonic at the photospheric level. To understand the driving mechanism, we investigated the acceleration by the Lorentz force and the pressure gradient force. Considering that the PIL is roughly straight and almost parallel to the unit vector $\hat{\bm{e}}=(-1/\sqrt{2},1/\sqrt{2})$, we made the dot products of the two force vectors and this unit vector to see the acceleration in the direction of the PIL: $f_{{\rm L,PIL}}\equiv(\bm{J}\times \bm{B})\cdot \hat{\bm{e}}$ and $f_{{\rm P,PIL}}\equiv(-\nabla p)\cdot  \hat{\bm{e}}$, respectively. Figure~\ref{fig:fdote} displays the relation between the two forces and horizontal flows. The color indicates $f_{{\rm L,PIL}}$ (Top) and $f_{{\rm P,PIL}}$ (Bottom). Note that positive (negative) $f_{{\rm L,PIL}}$ accelerates plasma in the upper-left (lower-right) direction. The same is true for $f_{{\rm P,PIL}}$. The horizontal velocity vector shows that the converging flows drastically change their direction to the direction of the PIL at the edges of the middle pair. From the figure, it is found that the shear flows along the PIL are driven by the Lorentz force. The pressure gradient force has no significant contribution (this implies that the hydrodynamic baroclinic vorticity generation ($-\nabla(1/\rho)\times \nabla p$) can be neglected). The acceleration regions coincide with the strong magnetic pressure regions at the edges of the middle pair, where the plasma $\beta$ is less than unity (see Figure~\ref{fig:pressure_balance_PIL}). The direction of the shear flow is determined by the expansion of each magnetic arcade: the magnetic arcade with the positive (negative) main polarity drives the shear flows in the positive (negative) $x$-direction.

We briefly compare our simulation with an observational example of $\delta$-spot regions. We focus on Active Region NOAA 11429 that was one of the most violent $\delta$-spot regions in the solar cycle 24. The $\delta$-spot region appeared in March 2012, and produced three X-class flares. Because of the high activity, it have been drawing many authors' attention \citep[e.g.][]{petrie2012,shimizu2014}. Figure~\ref{fig:obs} displays snapshots of the active regions and snapshots of the magnetogram obtained from our simulation (note that the sign of $B_z$ of the magnetogram from the simulation is reversed just for better comparison). The continuum and magnetogram images were taken by the Helioseismic and Magnetic Imager (HMI: \citet{schou2012}) on board the Solar Dynamics Observatory. The active region appeared in the northern hemisphere, and did not follow the Joy's Law. The mean force-free parameter $\langle \alpha \rangle$ of this region calculated in SHARP (Spaceweather HMI Active Region Patch) data was negative, which means that the region obeyed the helicity hemispheric rule. This is consistent with the apparent magnetic structure: the largest positive and negative spots show a left-handed twist. Our simulation reproduced some observed features of this $\delta$-spot region. The observed $\delta$-spot region has a narrow complex polarity pair between the main polarities (indicated by the arrow in the figure), and therefore it had a complex quadrupole structure. In addition, the middle pair appeared after the emergence of the main polarities. These observed features are also found in our simulation.

\section{DISCUSSION}\label{sec:discussion}
We have examined the evolution of the emergence of a kinked flux tube from the interior of the Sun to the corona using a 3D MHD simulation: we studied the rising of the kinked tube in the interior, emergence into the corona, and evolution of the coronal magnetic field. On the basis of the results, we discuss the kinked tube emergence scenario as a possible origin of the $\delta$-spot regions.

From the simulation, unlike the previous expectations based on the bodily emergence of a knotted tube \citep{fan1999,linton1999}, we found that the kinked tube can naturally form a complex quadrupole structure at the photospheric level. The appearance of the complex quadrupole structure has not been pointed out by the previous studies. This simulation is of much higher resolution compared to the previous study \citep{matsumoto1998}, and therefore we could find the development of the middle pair and the quadrupole morphology. The main magnetic polarities appeared first, and later another pair of magnetic polarities is formed between it as a result of the submergence of emerged magnetic fields. This leads to the formation of a quadrupole structure and two coronal magnetic arcades. 

\begin{figure}
\epsscale{.70}
\plotone{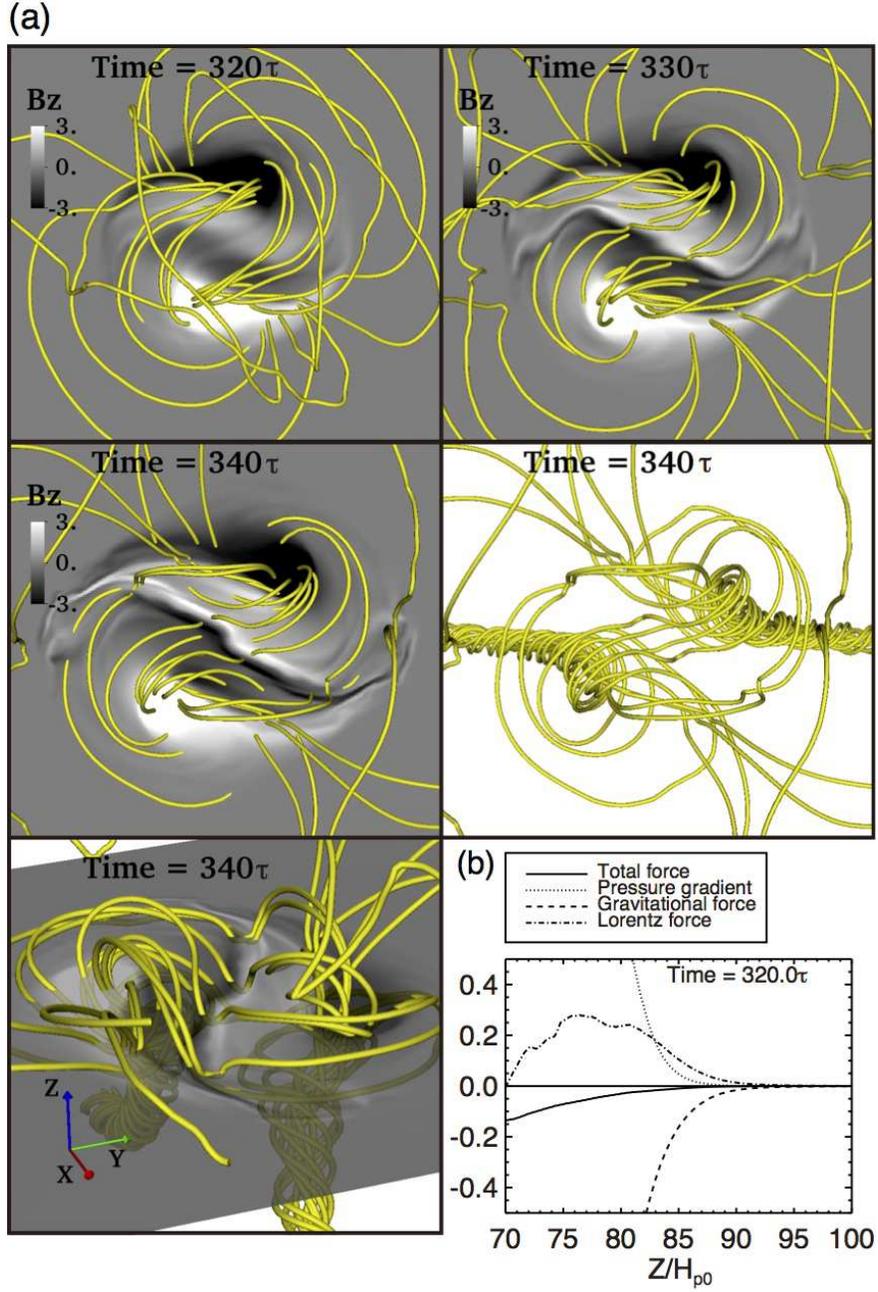}
\caption{Formation of the middle pair. (a) Temporal evolution of the 3D magnetic field (solid lines) and magnetogram $B_z$. The last two panels shows the magnetic structure below the photosphere and the dip of emerged magnetic fields at the time $t=340\tau$. (b) Profiles of the vertical forces measured at the center $(x,y)=(0,0)$. The total force (solid), pressure gradient force (dotted), gravitational force (dashed), and Lorentz force (dashed-dotted) are shown. Note that the sign of the total vertical force is negative near the photosphere, meaning that the downward gravitational force is the most dominant force.\label{fig:sinking3D}}
\end{figure}

\begin{figure}
\epsscale{.80}
\plotone{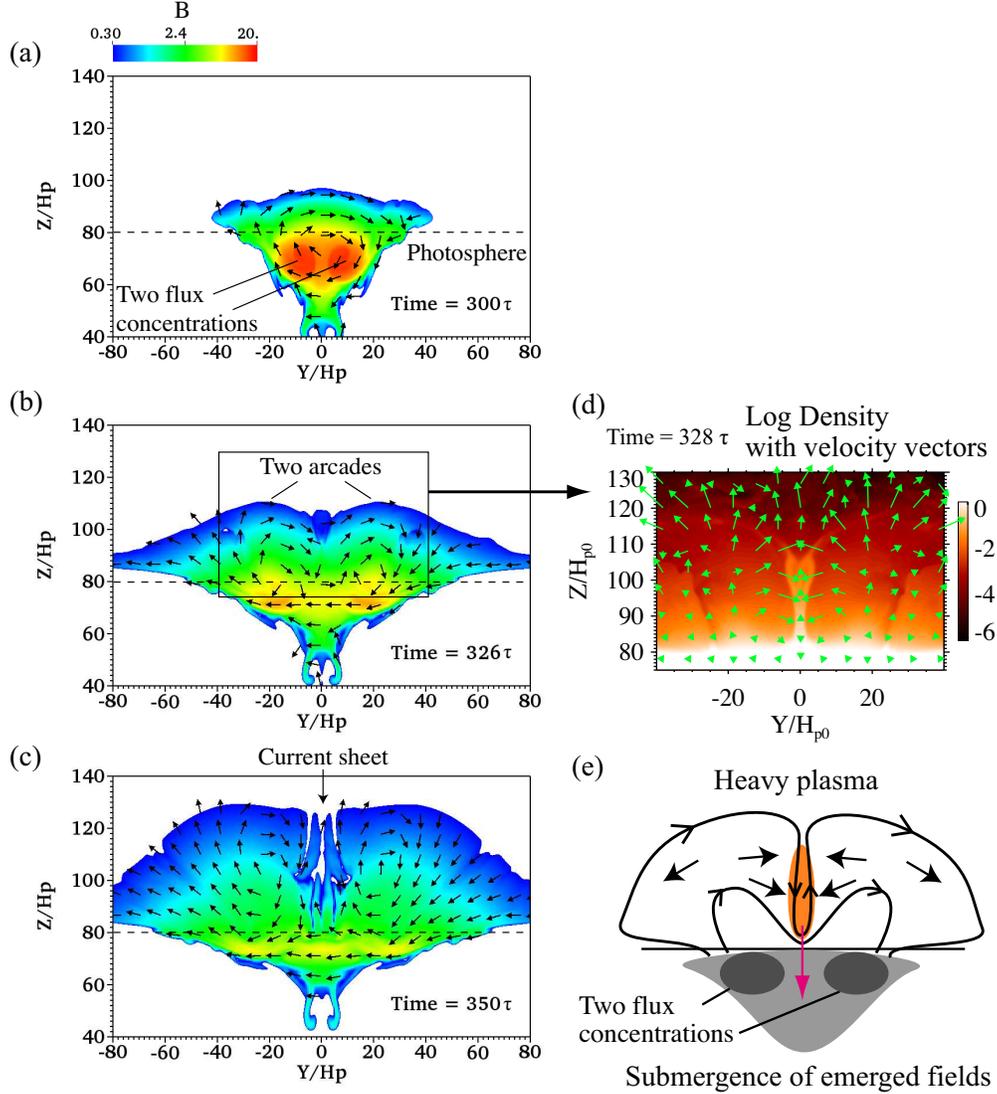}
\caption{(a-c): Time sequence of the magnetic field distribution on the $y$-$z$ plane at $x=0$. The arrows denote the direction of the magnetic field projected onto this plane (note that the size of the arrows does not represent the magnetic field strength). (d): The density distribution with velocity vectors on this plane. A high-density region is formed between the two arcades. (e): A schematic diagram to describe the submergence process. Plasma is accumulated in the middle of this region, and the heavy portion of the emerged fields submerges. \label{fig:2dcut}}
\end{figure}

\begin{figure}
\epsscale{.80}
\plotone{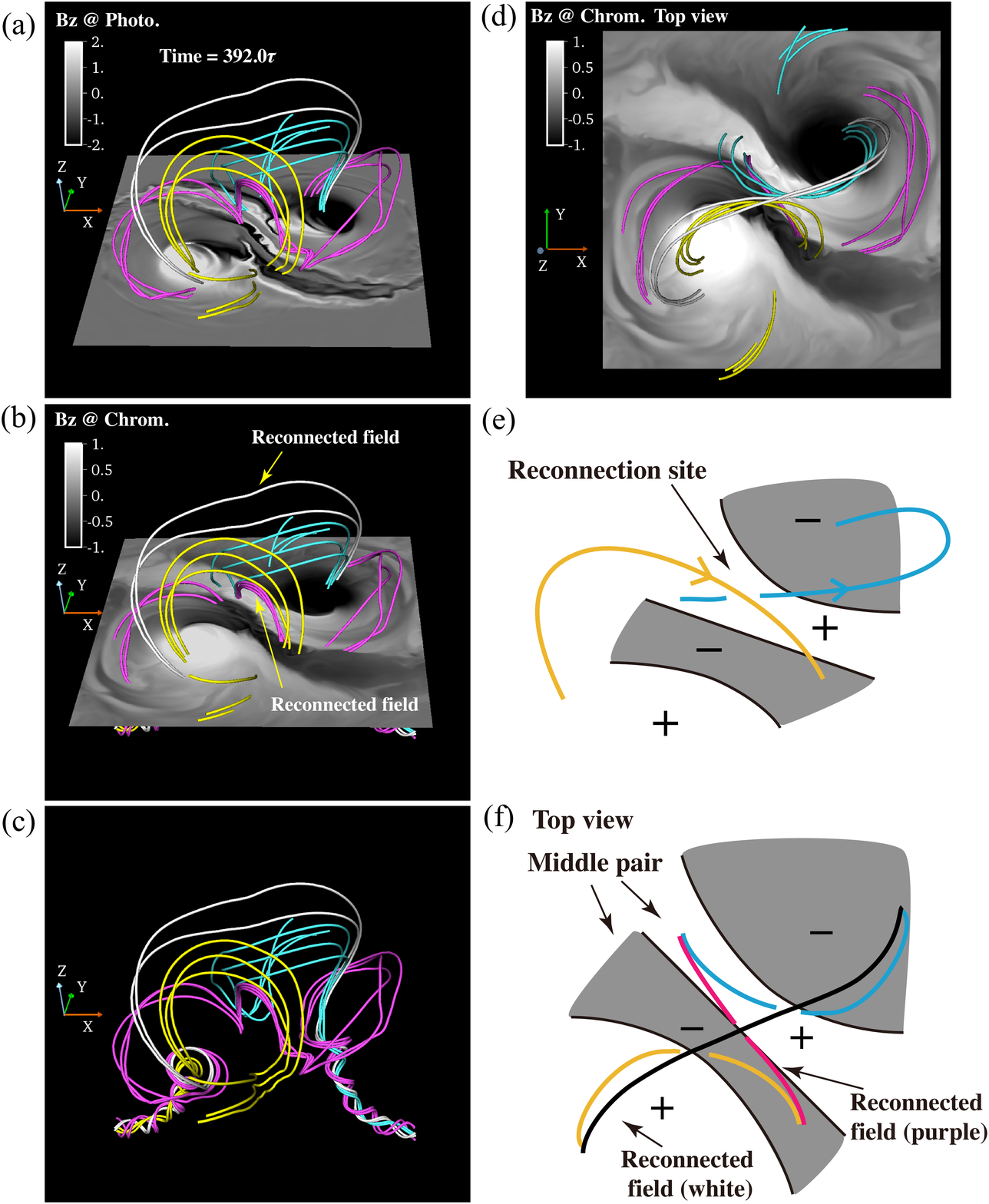}
\caption{3D Magnetic structure and photospheric and chromospheric line-of-sight magnetic fields $B_z$ at the time $392\tau$. The yellow and blue field lines denote field lines passing by the current sheet between the two arcades. The white field lines denote field lines enveloping the two arcades. The purple and white field lines denote field lines created by reconnection between the blue and yellow magnetic loops. (a-c): Bird's eye view. (d): Top view. (e): A schematic diagram of the merging field lines. (f): A schematic diagram of the magnetic field structure shown in the panel~(d). Note that the purple reconnected field lines are almost parallel to the neutral line at the middle. \label{fig:coronal_field}}
\end{figure}

Owing to the emergence of the two strong flux concentrations formed just below the photosphere, and the comparatively weaker twist at the middle between the two concentrations, the top heavy layer above the emerging fields is driven to accumulate plasma at a central dip, which results in the submergence of the heavy plasma and the formation of the quadrupole structure. A schematic diagram of the formation of a quadrupole photospheric structure and the two-arcade system is illustrated in Figure~\ref{fig:schematic_pic}. The kinking of the tube reduces the magnetic twist of the tube at the apex, and the magnetic field at the apex becomes more unidirectional (see the top panel of Figure~\ref{fig:schematic_pic} and Figure~\ref{fig:sinking3D}). As a result, the magnetic tension force to retain the coherency of the flux bundle decreases locally. Owing to the small tension force, the horizontal expansion of the apex of the tube is promoted. The magnetic field strength and the upward Lorentz force becomes weaker, causing the submergence of emerged fields. The bottom panel of the figure describes the flux emergence in the $y$-$z$ plane. The two flux concentrations on this plane are formed by the kinking of the axis (they correspond to the two parent flux tubes of the main spots). Two arcades develop from the two flux concentrations with a portion of the emerged fields undergoing submergence, which leads to the formation of a current sheet between them (see also Figure~\ref{fig:2dcut} and \ref{fig:coronal_field}). Finally magnetic reconnection takes place there to form a sheared magnetic arcade above the polarity inversion line.

\begin{figure}
\epsscale{.90}
\plotone{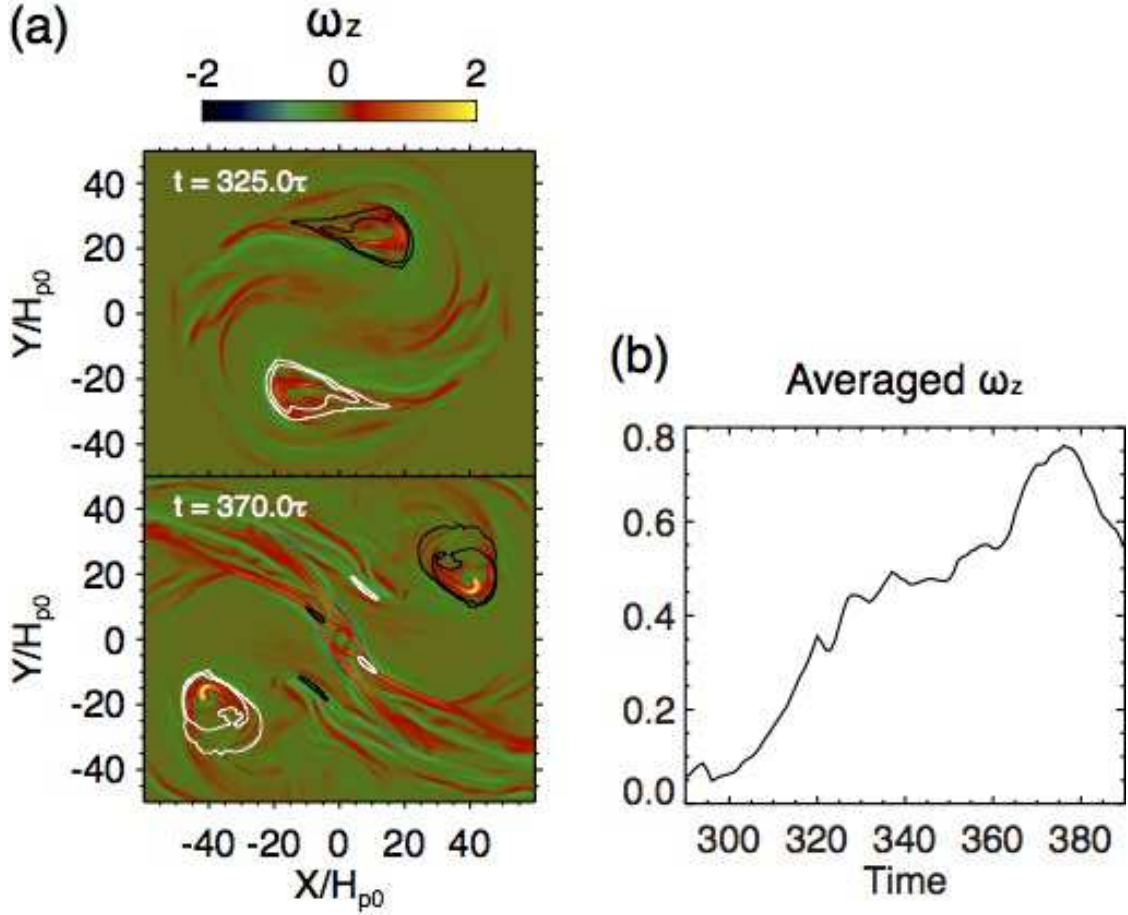}
\caption{(a) Snapshots of the $z$-component of the vorticity $\omega_z$ at the photospheric level. White and black contours denote positive $(3B_0,4B_0)$ and negative $(-4B_0,-3B_0)$ $B_z$, respectively. Note that the main polarities show counterclockwise motion. (b) Temporal evolution of the averaged $\omega_z$ over the area where $|B_z|$ is above 75\% of the peak  $|B_z|$. \label{fig:wz_photo}}
\end{figure}

\begin{figure}
\epsscale{.70}
\plotone{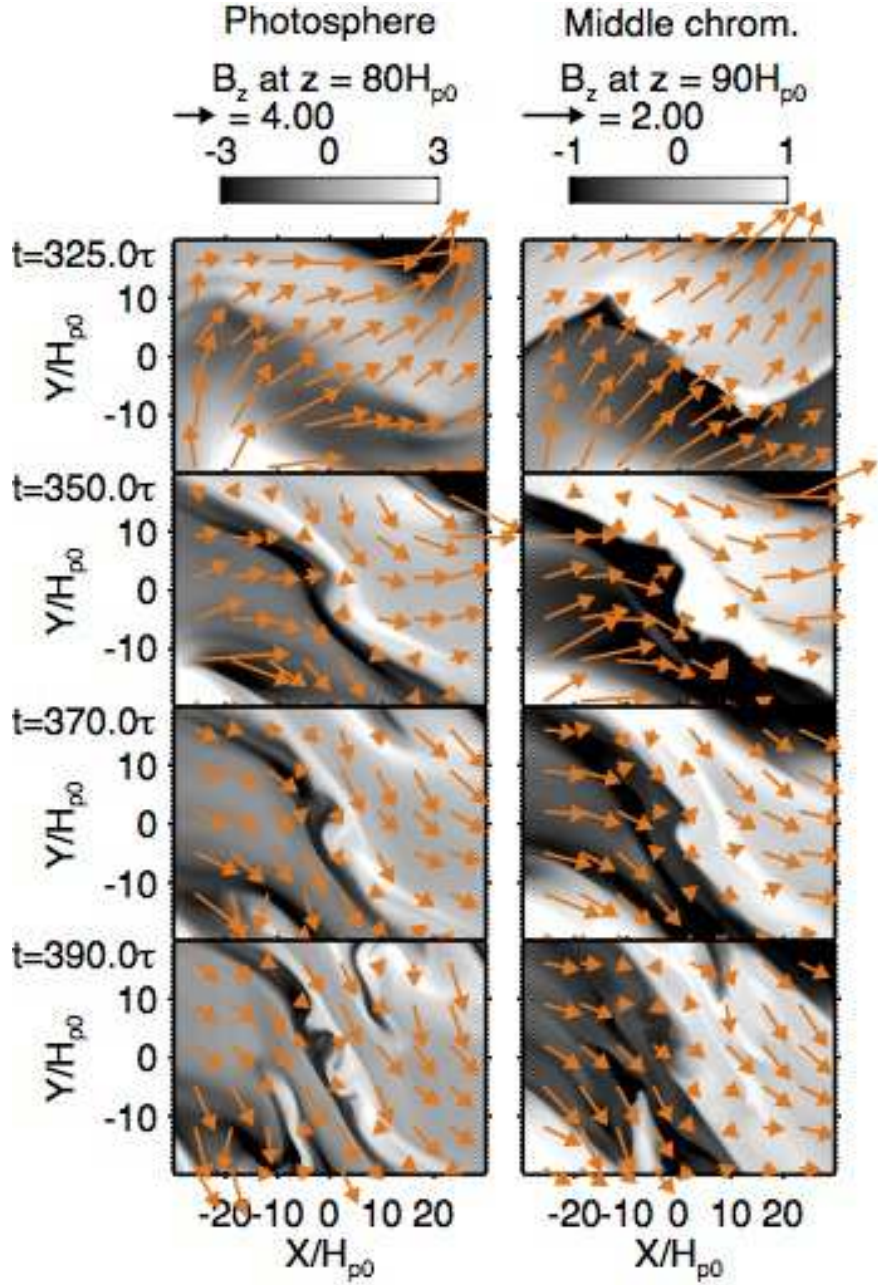}
\caption{Temporal evolution of the vector magnetogram of the photosphere and middle chromosphere. Arrows show the horizontal magnetic field vectors. The horizontal magnetic field becomes more parallel to the polarity inversion line (PIL) at those heights as time progresses. Note that the horizontal magnetic field strength is small on the PIL.  See also Figure~\ref{fig:wz_photo} as for magnetic shear development). \label{fig:bz_bhvec_multi_height}}
\end{figure}

\begin{figure}
\epsscale{.80}
\plotone{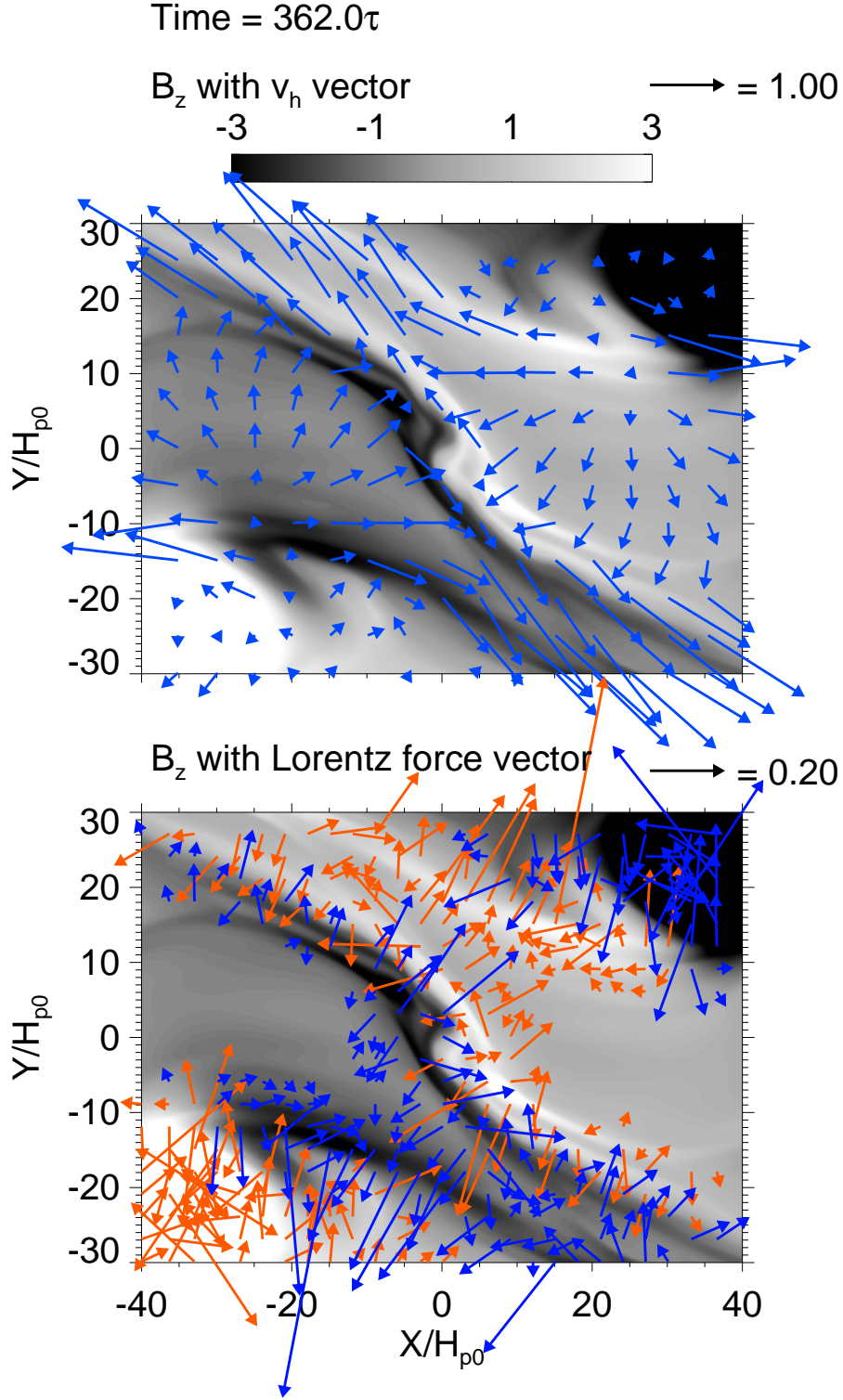}
\caption{Magnetogram ($B_z$) with horizontal velocity vectors (Top) and with Lorentz force vectors (Bottom) at time $t=362\tau$. In the bottom panel, red and blue arrows show the Lorentz force in positive and negative polarities, respectively. The velocity and Lorentz force are normalized by $C_{s0}$ and $B_0^2/H_{p0}$, respectively.\label{fig:bz_vhvec_lorentz}}
\end{figure}

\begin{figure}
\epsscale{.80}
\plotone{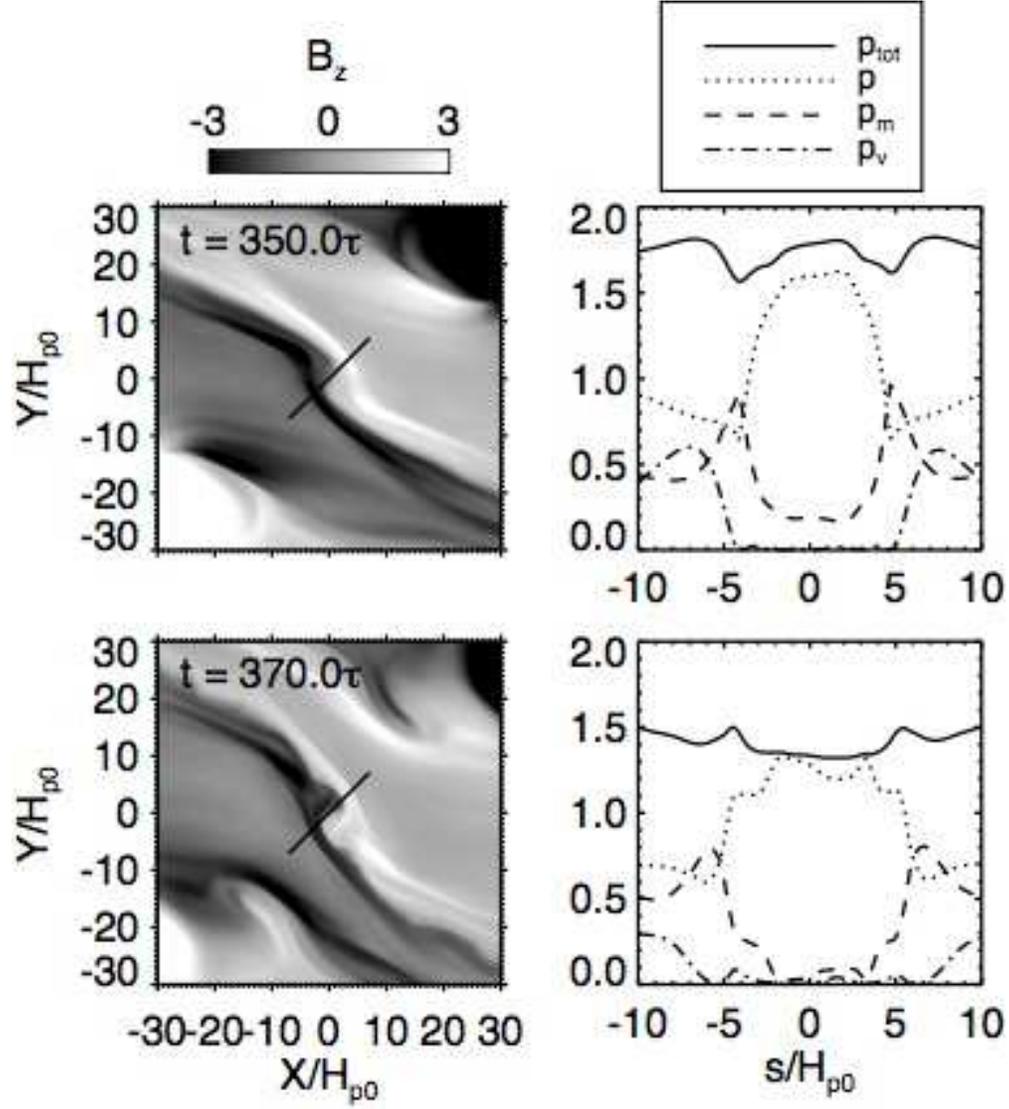}
\caption{Profiles of the total pressure $p_{\rm tot}$ (solid), gas pressure $p$ (dotted), magnetic pressure $p_m$ (dashed), and dynamic pressure $p_v$ (dashed-dotted) across the polarity inversion line of the middle pair. The profiles are measured along the slit shown in the $B_z$ maps. The velocity component parallel to the slit is used to calculate $p_v$. \label{fig:pressure_balance_PIL}}
\end{figure}

\begin{figure}
\epsscale{.90}
\plotone{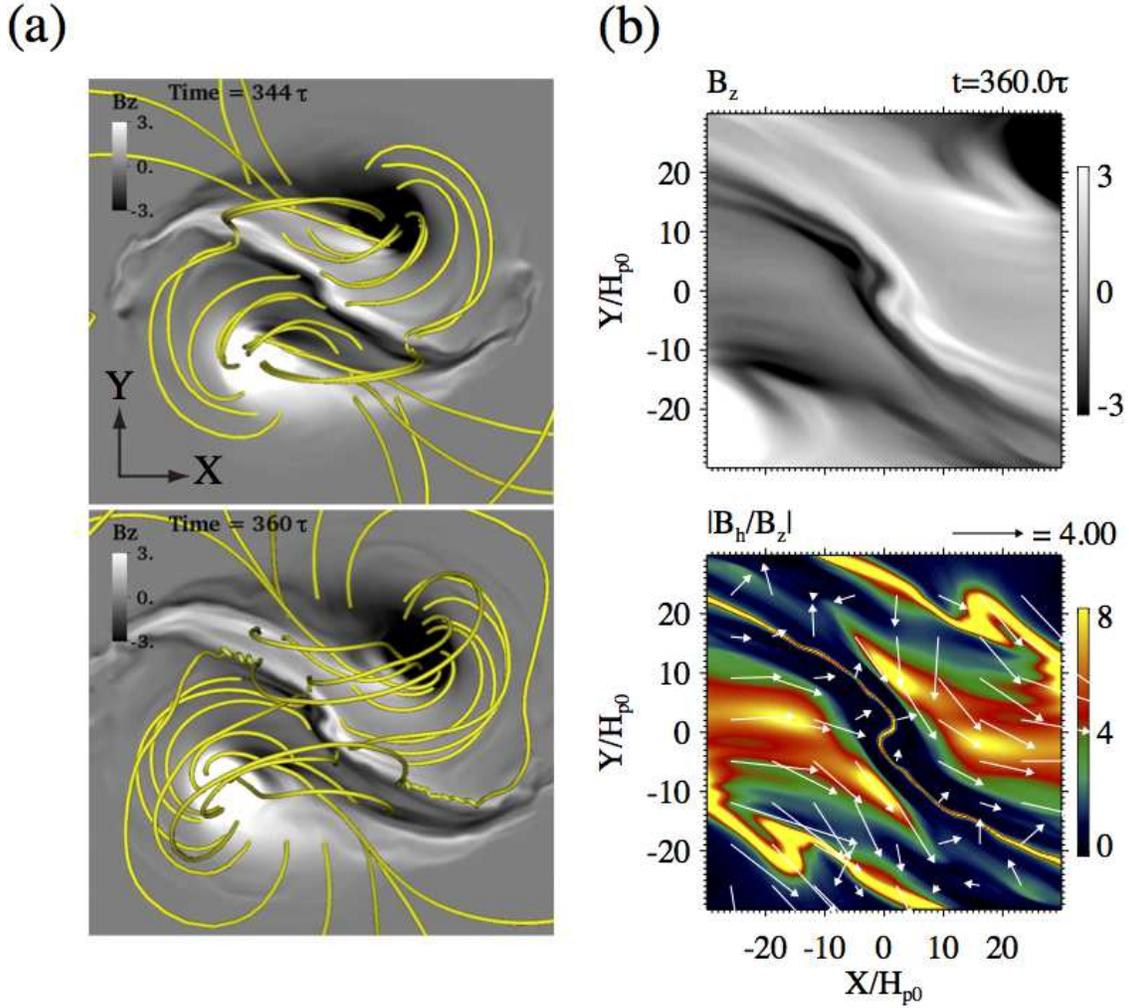}
\caption{Development of the horizontal magnetic field just near the polarity inversion line of the middle pair. (a) Snapshots of 3D magnetic field lines with magnetogram. (b) Enlarged images of the magnetogram $B_z$ and the ratio $|B_h/B_z|$ of the middle pair. The arrows indicate the horizontal magnetic field vector $\bm{B}_h=(B_x,B_y)$. \label{fig:shear_evo3D}}
\end{figure}

\begin{figure}
\epsscale{.80}
\plotone{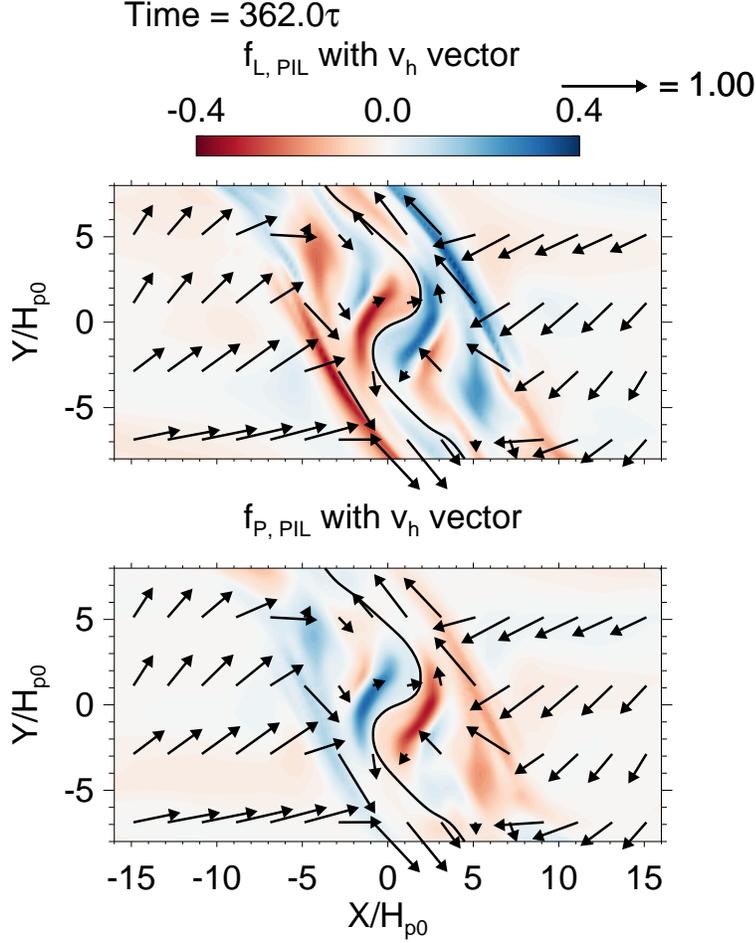}
\caption{The Lorentz force drives flows along the polarity inversion line (PIL) of the middle pair. The color indicates $f_{{\rm L,PIL}}=(\bm{J}\times \bm{B})\cdot \hat{\bm{e}}$ (Top) and $f_{{\rm P,PIL}}=(-\nabla p)\cdot  \hat{\bm{e}}$ (Bottom), where $\hat{\bm{e}}=(-1/\sqrt{2},1/\sqrt{2})$ is a unit vector almost parallel to the PIL. Note that positive (negative) $f_{{\rm L,PIL}}$ accelerates plasma in the upper-left (lower-right) direction. The same is true for $f_{{\rm P,PIL}}$. The arrows denote the horizontal velocity, and the solid lines indicate the PIL ($B_z=0$). The velocity is normalized by $C_{s0}$. \label{fig:fdote}}
\end{figure}

\begin{figure}
\epsscale{.90}
\plotone{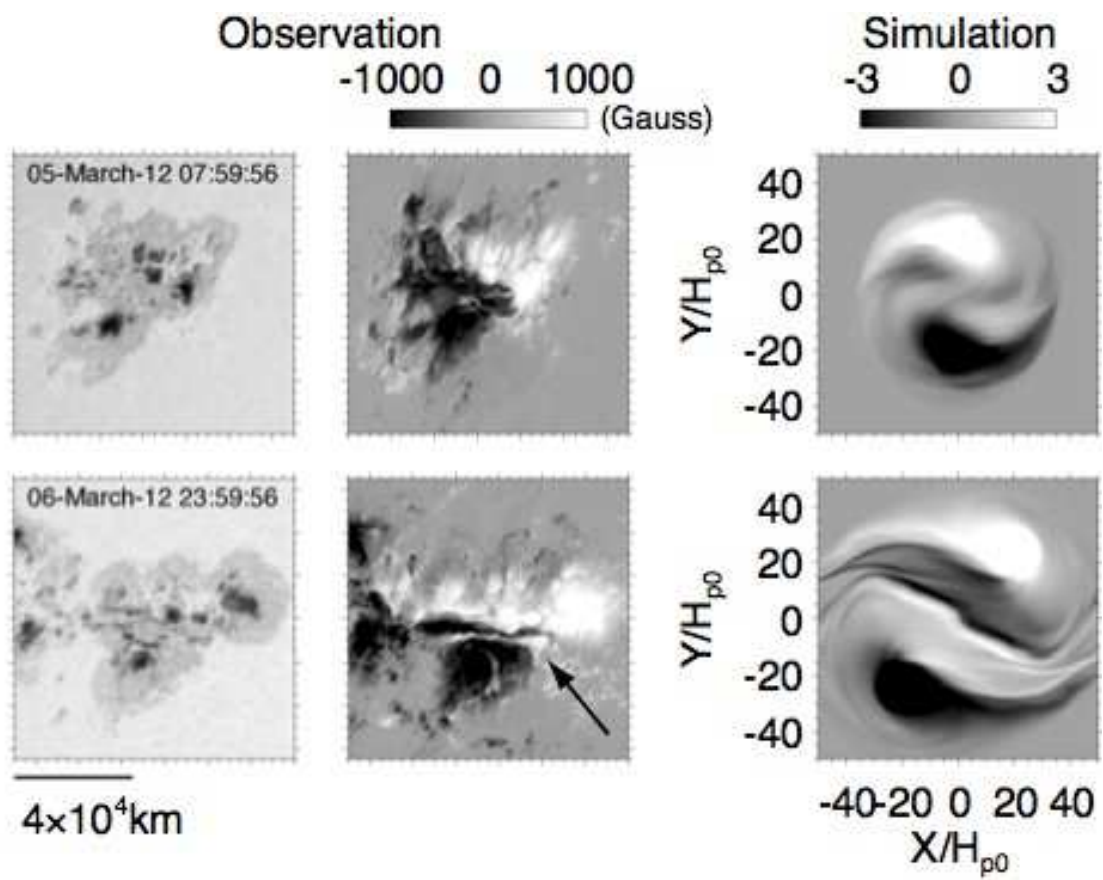}
\caption{Comparison of the $\delta$-spot region NOAA 11429 and our simulation. The left and middle panels show the continuum and magnetogram images. The middle polarity pair between the largest opposite polarities is indicated by the arrow. The right panels show the snapshots of the magnetogram from our simulation at the time $t=310\tau$ and $335\tau$. Note that the sign of $B_z$ of the magnetogram from the simulation is reversed just for better comparison. This does not change its twist handedness. \label{fig:obs}}
\end{figure}

\begin{figure}
\epsscale{.70}
\plotone{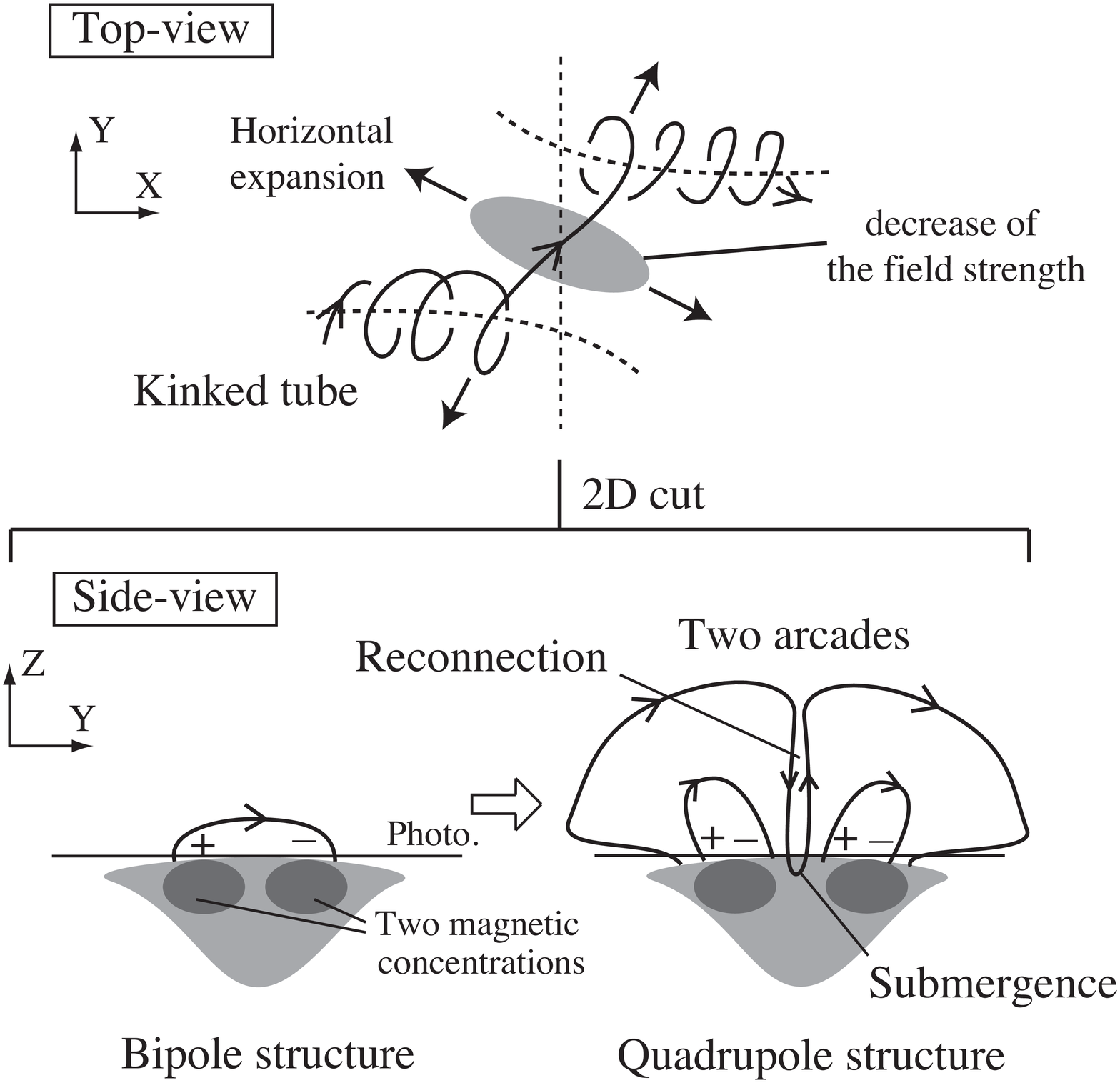}
\caption{Schematic diagram of the formation of a quadrupole photospheric structure. Top: the horizontal expansion of the kinked flux tube before the emergence. See also Figure~\ref{fig:rising_tube} and \ref{fig:sinking3D}. Bottom: 2D cut view of the expanding two magnetic arcades. The two magnetic concentrations below the photosphere correspond to the two parent flux tubes. The plus and minus signs denote the positive and negative polarities at the photosphere, respectively. See also Figure~\ref{fig:2dcut} and \ref{fig:coronal_field}.\label{fig:schematic_pic}}
\end{figure}

The kinked part, which is the most buoyant part, is subject to the kinking and strong horizontal expansion. As a result, the twist and field strength are largely reduced. This means that some of the free magnetic energy of this segment is already released before the emergence. However, the twist and field strength of the parent flux tubes remain strong, which means that a large amount of the free energy is stored there. The free energy is injected in the form of the vortical motion of the main polarities after the emergence, and the less-sheared coronal field is twisted up eventually (Figure~\ref{fig:wz_photo}, \ref{fig:bz_bhvec_multi_height}, and \ref{fig:shear_evo3D}). Also, the highly nonuniform distribution of the twist and field strength along the tube will explain that the middle pair has the less-coherent photospheric structure and the main pair has a strong twist with a coherent structure.

\citet{fan1999} and \citet{linton1999} speculated the origin of the sheared field above a polarity inversion line (PIL) by simply taking horizontal cuts of the kinked tubes approaching the surface, and argued that the shear can be introduced by the kinking of a tube. In this study, on the contrary, the magnetic shear along the PIL is weak in the early phase of the emergence, which means that the direct emergence does not provide a strong sheared field above the photosphere. The strong magnetic shear is built-up by the vortical motion of the main polarities.

As another possible origin of $\delta$-spot regions, \citet{toriumi2014} and \citet{fang2015} proposed a simple scenario in which a single twisted flux tube with two buoyant segments emerges (here we refer their model as the``multi buoyant segment model, MBS model"). It was found that this model successfully accounts for many observational characteristics of the active region with a quadrupole structure NOAA~11158. Here we compare the kinked tube model (KT model) and their MBS model. In both models, a quadrupole photospheric structure can be formed, and the middle pair is locked by the Lorentz force. In addition, the middle pair in both models are magnetically connected below the surface. However, we see a difference in the photospheric structure between the two models. The middle pair found in KT model shows a narrow and incoherent structure, but every spot in the quadrupole found in MBS model can have a coherent structure. In KT model, we found the flux imbalance between the main pair and the middle pair. However, the flux imbalance in MBS model is less prominent. Photospheric polarity motion and magnetic reconnection are important to form a sheared field on and near the PIL in both models. The magnetic shear in KT model is introduced mainly by the vortical motion of the main polarities. The shear in MBS model is introduced by the stretching of the reconnected field by the horizontal spot motion.

We observe supersonic shear flows along the PIL of the middle pair. We identified that the driving force is the Lorentz force, as found in previous studies \citep{manchester2001,fan2001}. The flows are driven by the Lorentz force at the edges of the middle pair, where the plasma $\beta$ is less than unity. The strong magnetic field at the edges is resulted from the squeezing by the high gas pressure and the dynamic pressure of the converging flows. The converging flows are driven by the expansion of two magnetic arcades, and the direction of the shear flow is also determined by the expansion. Therefore, the generation of the shear flows is a direct consequence of the development of the two arcades.

We briefly compared our simulation with a $\delta$-spot region NOAA 11429, and found similarities in the behavior of the photospheric magnetic structures. The $\delta$-spot region has a narrow complex polarity pair between the main polarities. In addition, the middle pair appeared after the emergence of the main polarities. On the basis of our simulation, we expect that the spots in the $\delta$-spot region are magnetically connected below the photosphere, which is important for keeping $\delta$-spot regions compact. We note that we have to be careful to compare our simulations to observations: the total amount of the magnetic flux of a main spot is only $3\times10^{20}$~Mx (smaller than a typical value, $\sim10^{22}$~Mx), and the convection and radiative cooling are not included. Because of the lack of the radiative cooling, the penumbra that is important to define the $\delta$-spot regions cannot be formed in the simulation. In this study, we hypothesize that the polarities of the middle pair is locked closely enough to be surrounded by a common penumbra. To confirm the speculation, more realistic simulations and detailed comparison with observations are necessary.

\citet{takizawa2015} performed a statistical study, and found that flare-productive $\delta$-spot regions tend to have a quadrupole structure and are likely to be formed by the emergence of a singly connected structure. In addition, they clarified that the flare activity is highly correlated with the magnetic complexity. Our simulation showed the formation of a complex quadruple structure from a single flux tube. Considering this, we conjecture that our results may give a general picture of the formation of the highly flare-productive $\delta$-spot regions.

It has been argued that magnetic reconnection between sheared fields is important for the onset of solar flares \citep[e.g.][]{moore2001}. It has been also discussed that a quadrupole structure is preferable for the filament eruption \cite[e.g.][]{antiochos1999,hirose2001,yurchyshyn2006}. Our simulation shows that the coronal magnetic structure changes its topology via reconnection between sheared arcades in a quadrupole region, although the formation and eruption of a flux rope was not achieved probably due to a short-term calculation in a smaller domain size than the size of a real active region \citep[Figure~\ref{fig:coronal_field}. We note the similarity between the reconnection in our Figure~\ref{fig:coronal_field} and the reconnection in Figure~1 of][]{moore2001}. Considering this, our results could be relevant for understanding how the formation of $\delta$-spot regions with a quadrupole structure can lead to the onset of solar flares. For comprehensive understanding including the onset of flares, it is necessary to perform longer-term calculations in a larger calculation domain size.

\acknowledgements
ST acknowledges support by the Research Fellowship of the Japan Society for the Promotion of Science (JSPS). 
This work was supported by a Grant-in-Aid from the Ministry of Education, Culture, Sports, Science and Technology of Japan (No. 25287039).
This work was supported by the ``UCHUGAKU" project of the Unit of Synergetic Studies for Space, Kyoto University. 
Numerical computations were carried out on Cray XC30 at Center for Computational Astrophysics, National Astronomical Observatory of Japan.
The HMI data have been used courtesy of NASA/SDO and the HMI science teams. The National Center for Atmospheric Research is sponsored by the National Science Foundation. MCMC acknowledges support from NASA’s Heliophysics Grand Challenges Research grant (NNX14AI14G).

\begin{table}
\begin{center}
\caption{Normalization units\label{tab:units}}
\begin{tabular}{crr}
\tableline\tableline
Quantity & Unit & Value\\
\tableline
Length & $H_{p0}$ & 170~km\\
Velocity & $C_{s0}=\left[ \gamma(k_B/m)T_{0} \right]^{1/2}$ & 6.8~km~s$^{-1}$ \\
Time & $\tau=H_p/C_{s0} $ & 25~s \\
Temperature & $T_{0}=T_{ph}$ & 5,600~K\\
Density & $\rho_0=\rho_{ph}$ & $1.4\times10^{-7}$~g~cm~$^{-3}$\\
Pressure & $\gamma (k_B/m)\rho_0 T_0$ & $6.3\times 10^{4}$~dyn~cm$^{-2}$\\
Magnetic field strength & $B_{0}$ & 250~G\\
\tableline
\end{tabular}
\end{center}
\end{table}

\end{document}